\documentclass[12pt,a4paper]{article}
\usepackage{amsmath,amstext,amssymb,amscd}

\input{tcilatex}
\begin{document}

\title{Targeting in Quantum Persuasion Problem}
\author{V. I. Danilov\thanks{%
CEMI Russion Academy of sciences, Moscow, vdanilov43@mail.ru} and A.
Lambert-Mogiliansky\thanks{%
Paris School of Economics, alambert@pse.ens.fr}\thanks{%
We would like to thank James Yearsley as well as an anonymous referee for
very valuable discussions and comments. }}
\date{April 20, 2018}
\maketitle

\begin{abstract}
In this paper we investigate the potential for persuasion arising from the
quantum indeterminacy of a decision-maker's beliefs, a feature\textbf{\ }%
that has been proposed as a formal expression of well-known cognitive
limitations.\textbf{\ }We focus on a situation where an agent called Sender
only has few opportunities to influence the decision-maker called Receiver.
We do not address the full persuasion problem but restrict attention to a
simpler one that we call targeting, i.e. inducing a specific belief state.
The analysis is developed within the frame of a $n-$dimensional Hilbert
space model. We find that when the prior is known, Sender can induce a
targeted belief with a probability of at least $1/n$ when using two
sequential\textbf{\ }measurements$.$ This figure climbs to 1/2 when both the
target and the belief are known pure states. A main insight from the
analysis is that a well-designed strategy of distraction can be used as a
first step to confuse Receiver. We thus find that distraction rather than
the provision of relevant arguments is an effective means to achieve
persuasion. We provide an example from political decision-making.
\end{abstract}

\section{Introduction}

In this paper we build further on a theoretical result \cite{TCS} that
establishes the power of unconstrained belief manipulation or persuasion in
the context of non-classical (quantum) uncertainty. That result was a first
step in the development of Quantum persuasion theory in the spirit of
`Bayesian Persuasion' by Kamenica and Gentskow \cite{KG11}, hereafter KG.
Bayesian persuasion should not be confused with the theory of persuasion in
communication games where an informed party (Sender) with certifiable
information chooses what pieces of information to reveal to an uninformed
decision-maker (Receiver). The subject matter of Bayesian persuasion is the
choice by Sender of an `information structure' (or measurement). In
particular in KG, Sender is not better informed so they are not concerned
with strategic revelation (or concealment) of information. The only choice
variable of Sender is the measurement itself which is performed so the
resulting signal is revealed publicly. And the question is how much can be
achieved in terms of modifying a rational Receiver's belief by an
appropriate choice of measurement. The ultimate goal is to influence
Receiver's decision to act which depends on her beliefs about the world.

KG investigate the issue in the classical uncertainty framework. Receiver's
beliefs are expressed as a probability distribution over the set of states
of the world and updating follows Bayes rule. However as amply documented
the functioning of the mind is more complex and often people do not follow
Bayes rule. Cognitive sciences propose various alternatives to Bayesianism.
One avenue of research within cognitive sciences appeals to the formalism of
quantum mechanics. A reason is that QM has properties that reminds of the
paradoxical phenomena exhibited in human cognition. But the motivation for
turning to QM is deeper. Indeed, the two fields share fundamental common
features namely that the object of investigation and the process of
investigation cannot always be separated. This similarity was already put
forward by the fathers of quantum mechanics. Its mathematical formalism was
developed to respond to that epistemological challenge (see \cite{Bit08}).
In addition, quantum cognition has been successful in explaining a wide
variety of behavioral phenomena (for a survey see Bruza and Busemeyer \cite%
{Bubu12}). Finally, there exists by now a fully developed decision theory
relying on the principle of quantum cognition (see \cite{danalm10, danalmver}%
). Therefore in the following we shall use the Hilbert space model of QM to
represent the beliefs of an individual and capture the impact of new
information on those beliefs. Clearly, the mind is likely to be even more
complex than a quantum system but our view is that the quantum cognitive
approach already delivers interesting new insights in particular with
respect to persuasion.

In quantum cognition, we distinguish between the "world" and the
decision-maker's mental representation of it which is the basis for her
decision. This representation of the world is modeled as a quantum-like
system and characterized by a \textit{cognitive} state. The decision
relevant uncertainty is of non-classical (quantum) nature. As argued in
Dubois et al. (\cite{Dualm16}) this modeling approach allows capturing
widespread cognitive limitations that people exhibit when constructing a
mental representation of a `complex' alternative. The key quantum property
that we use is that some characteristics (or properties) of a complex mental
object may be `Bohr complementary' that is incompatible in the
decision-maker's mind: they cannot have definite value simultaneously. A
central implication is that measurements (new information) modifies the
cognitive state in a non-Bayesian well-defined manner.

The kind of situations that we have in mind can be illustrated by the
following story. A member of parliament (MP) is considering voting for a law
to introduce a state of emergency in order to fight against terrorism. The
terror threat can be either severe or moderate. If the threat is perceived
as severe enough by our MP, she will support the law but not otherwise.
Initially, she believes that the threat is severe with a high probability,
so she would support the law. But a civil liberty activist wants her to vote
down the law. Instead of trying to bring forward arguments about the actual
threat, he brings up another topic: the EU process of decision-making and
its intrusive regulatory policies. In particular, he proposes to find out
whether the EU is regulating the weight of cucumbers - our MP represents
farm owners. EU spent many months reflecting on cucumbers and ended up with
a strict regulation which is broadly perceived as nonsensical. After this
intermezzo, they return to the emergency issue and our MP does not feel as
convinced as before and chooses to vote down the law. In section 3 we show
how this story can be formalized and explained in terms of quantum
persuasion.

{Our results }in{\  \cite{TCS} shows that with an unbounded sequence of
measurements, the belief state of a quantum-like Receiver can be brought
into any desirable state starting from any initial unknown state.
Interestingly, the idea of turning to a sequence of measurements never
arises in the classical context because any sequence of measurements can be
merged into a single one (possibly very hard to implement in practice
though). In the non-classical context it is generally not possible to merge
several measurements }into one direct (projective) measurement unless they
are compatible. But that framework allows accounting for incompatible
measurements performed in a sequence.{\ This incompatibility is an
expression of the fact that measurements modify the system. How new
information modifies the cognitive state is established in \cite{danalmver}.%
\footnote{%
Updating follows L\H{u}ders' rule which generalizes Bayes' rule to the
quantum context.}\ In order to extract results of more practical
significance, we here investigate the polar case when only a limited
sequence of measurement is feasible; more precisely we shall focus on the
case of one or two measurements.}

{Because of the highly non-linear structure of the problem, }we investigate{%
\textbf{\ }a simpler task that we call \textquotedblleft
targeting\textquotedblright . The object of targeting is the transition of a
belief state into another specified target state. We further rconfine
attention to projective measurements and combinations of those. This is
because our focus is on belief manipulation or state transition.\footnote{%
The most general type of measurements called POVM (positive operator valued
measurements) do not allow addressing state transition without additional
restrictions.}}

Our main result (Theorem 2) shows that if the initial belief state (prior)
is known, any target state can be reached with a probability of at least $%
1/n $ (where $n$ is the dimension of the corresponding Hilbert space). This
conclusion is also true for unknown prior when the target is a pure state.
When both the prior and the target are pure states, the probability for
successful manipulation increases to at least 1/2.

The path of measurements is of interest on its own. The second measurement
fully determines Receiver's decision: we ask Receiver to determine herself
with respect to the decision relevant uncertainty. For instance: do you
think this used smartphone is worth more than 400 euro or less when the
selling price is 400 euro. Clearly, if she thinks it is worth less than 400
euros, it is equivalent to a decision not to buy. However in a non-classical
context, the elicitation (measurement)\ of the beliefs changes the state so
it is not equivalent to making the decision on the basis on the non elicited
(unmeasured) belief.\footnote{%
The impact of eliciting beliefs on decision-making has been exhibited in
experimental set-ups see, e.g. \cite{Erev93}.} What we find most interesting
is the characterization of the first measurement which aims at creating a
state of uniform uncertainty. We show in Theorem 1 that Sender can always
transform known Receiver's prior into uniform uncertainty which explains
appearance of the probability $1/n$. {This result is in clear contrast with
the classical context where posteriors are constrained by Bayesian
plausibility.\footnote{%
Bayesian plausibility entails that that the expected posteriors must equal
the priors. }}

One interpretation is that the power of quantum persuasion is related to the
possibility of creating confusion in Receiver's mind. The strategy involves
exploiting the incompatibility of perspectives (non-commutativity of
measurements). Basically, this is a strategy of `calculated diversion' to
instigate confusion. Once Receiver is confused, it is much easier to
persuade her to do what Sender wants. \medskip

{The seminal paper by Kamenica and Gentskow \cite{KG11} analyzes the general
problem of persuading a rational agent by controlling her information
environment. It was followed by a series of other papers addressing
competition in persuasion, costly persuasion, and most recently Bayesian
persuasion with multiple senders. It also gave rise to a number of
applications, see for instance \textquotedblleft Persuading
Voters\textquotedblright \ by Alonso Camara \cite{alonso16}. Our
contribution extends that literature with a theoretical development to
non-classical uncertainty. }It also relates to more applied issues. Akerlof
and Shiller \cite{AS} argue that manipulation is determinant to the
functioning of markets. They suggest that it goes well beyond Bayesian
updating: \textquotedblleft Just change people's focus and one can change
the decisions they make\textquotedblright \ (p.173). In this paper we show
that quantum persuasion is powerful even for short sequences of measurements
and that it delivers a mechanism where merely changing the focus of the mind
can significantly affect decision-making.

The paper proceeds as follows. In Section 2 we sketch the persuasion problem
in the classical setting and thereafter formulate the corresponding problem
in the quantum context. In Section 3 we analyze the targeting task and
derive our main results. We provide an illustration in terms of the example
presented in the Introduction. We conclude with some remarks.

\section{The model}

The persuasion problem has been formulated by Kamenica and Genskow (KG) \cite%
{KG11} as follows . There are two players called respectively Receiver and
Sender. Receiver chooses on action with uncertain consequences. To assess
the value of the different actions Receiver uses her beliefs about the state
of the world. Receiver's action also has consequences for Sender. Therefore
Sender may try to persuade Receiver so she chooses an action favorable to
him. For that purpose Sender's selects some information structure (or
measurement) that generates new information (or signal) about the relevant
uncertainty. Below we shortly remind of the formulation of the problem in
the classical setting (in its simplest form). Thereafter, we describe in
details the corresponding formulation in the context of non-classical
(quantum) uncertainty.

\subsection{The classical setting}

Classical, or Bayesian, persuasion has been well described by KG. The
uncertainty is formulated using a set $\Omega $\ of states of Nature. For
the sake of simplicity we assume that $\Omega $ is a finite set. An \emph{%
action } is a function $a:\Omega \rightarrow \mathbb{R}$; the number $%
a(\omega )$ is a utility of Receiver from action $a$ at the state $\omega $.
Receiver has to choose an action from some finite set of available actions.
Since she does not know the true state of the world, Receiver relies on her
belief to make her decision.

Receiver and Sender share a prior belief that is a probability distribution $%
\beta $\ on $\Omega \ $reflecting a common (objective) representation of the
world i.e., a collection of probabilities\textbf{\ }$\beta (\omega )\geq 0$%
\textbf{, }$\sum_{\omega }\beta (\omega )=1$\textbf{.} Let $\Delta (\Omega )$
denote the simplex of probability distributions on $\Omega $. The expected
utility of an action $a$ under a belief $\beta $ is $a(\beta )=\sum_{\omega
}a(\omega )\beta (\omega )$. Thus an action can be represented as an affine
function on $\Delta (\Omega )$. \ Given a belief $\beta $ Receiver chooses
an action $a^{\ast }$ that maximizes her expected utility $a^{\ast }(\beta )$%
.

Action $a$ brings Sender utility $u(a)$. (Here we assume that Sender's
utility only depends on the action chosen by Receiver and not on the state
of the Nature). Sender tries to influence receiver's choice by providing
some additional information. In order to obtain that information, Sender
performs some measurement (of Nature) and informs Receiver about the result
of the measurement.

An \textit{information structure} is a map $f:\Omega \rightarrow \Delta (S)$%
, where $S$ denotes the set of \emph{signals}. In state $\omega \in \Omega $
the device generates a (randomized) signal $f(\omega )\in \Delta (S)$. If we
write this more carefully an information structure is given by a family ($%
f_{s},\ s\in S)$ of non-negative function $f_{s}:\Omega \rightarrow \mathbf{%
R_{+}}$; $f_{s}(\omega )$ gives the probability of obtaining signal $s$ in a
state $\omega \in \Omega $. Of course, $\sum_{s}f_{s}$ must yield the unit
function $1_{\Omega }$ on $\Omega $.

When Receiver obtains a signal $s$, she, as a rational decision maker,
updates her initial belief (or prior) $\beta $ using Bayes' rule, i.e. she
forms the posterior $\beta _{s}\in \Delta \left( \Omega \right) $, given as $%
\beta _{s}(\omega )=f_{s}\left( \omega \right) \beta \left( \omega \right)
/p_{s}$, where $p_{s}=\sum_{\omega }\beta \left( \omega \right) f_{s}\left(
\omega \right) $ is the probability of signal $s$. As a consequence, she
chooses a new action $a^{\ast }(\beta _{s})$ and Sender receives utility $%
u\left( a^{\ast }\left( \beta _{s}\right) \right) $. When Sender uses such
an information structure, his expected utility is equal to $\sum
p_{s}u\left( a^{\ast }\left( \beta _{s}\right) \right) .$ KG provide a
characterization of the optimal measurement i.e., of the information
structure that maximizes Sender's expected utility. Bayesian plausibility, a
property that says that the expected posterior is equal to the prior, $\beta
=\sum_{s}p_{s}\beta _{s}$ plays a key role in their characterization.

\subsection{The quantum setting}

The description of a quantum system starts with the fixation of a Hilbert
space (in our case finite dimensional) over the field of complex or real
numbers. The choice of field does not affect our results so for the sake of
simplicity we choose to limit ourselves to real numbers (the complex case
obtains with minor changes). In this case $H$\ is simply an Euclidian space
equipped with a symmetric scalar product (.,.).

We shall be interested not so much in the Hilbert space $H$ as in operators
(that is linear maps from $H$ to $H$). For such an operator $A$ we denote by 
$A^{\ast }$ its conjugate which is defined by the following condition: $%
(v,Aw)=(A^{\ast }v,w)$ for all $u,w$ in $H$. Self-conjugate operators (for
which $A=A^{\ast })$ are called \textit{Hermitian}\footnote{%
More correctly we should speak here about symmetric operators; we use the
term Hermitian holding in mind the extension to the complex case.}.

We model actions as Hermetian operators.\textbf{\ }The utility of an action
for Receiver depends on her belief about the state of the relevant quantum
system, see \cite{danalmver}. We elaborate on this below, after introducing
two additional notions.\medskip

An Hermitian operator $A$ is \emph{non-negative} if $(x,Ax)\geq 0$ for any $%
x\in H$. For example, any \emph{projector} (that is a Hermitian operator $P$
with the property $PP=P$) is non-negative. Indeed, $(x,Px)=(x,PPx)=(Px,Px)%
\geq 0$. More generally, for any operator $C$ the operator $C^{\ast }C$ is
non-negative. \medskip

The notion of trace of operators is a central instrument in what follows. It
associates to an operator $A$ the number $\mathbf{Tr}(A)$ called its \emph{%
trace}. We collect in Proposition 1 below the properties of the trace that
we use throughout the paper.\medskip

\textbf{Proposition 1.} i) $\mathbf{Tr}(A)$ \emph{is linear over $A$;}

ii) $\mathbf{Tr}(AB)=\mathbf{Tr}(BA)$;

iii) \emph{If an operator is represented by a square matrix $A=(a_{ij})$,
then $\mathbf{Tr}(A)=a_{11}+...+a_{nn}$, that is $\mathbf{Tr}(A)\ $equals to
the sum of the diagonal terms of the matrix. }

iv) \emph{If $A$ is a non-negative operator then $\mathbf{Tr}(A)\geq 0$, and
equal to 0 only for }$A=0$.\medskip

\textbf{Definition}. A (cognitive) \emph{state} is a non-negative
(Hermitian) operator $S$ with $\mathbf{Tr}(S)=1$. \medskip

The set of states is denoted as $\mathbf{St}$. It is the quantum analog to
the classical set $\Delta (\Omega )$. Indeed the notion of cognitive state
reminds of the notion of belief in the classical context. The non-negativity
of operator $S$ is analogous to the non-negativity of a probability measure,
and the trace 1 to the sum of probabilities which equals 1. In a way similar
to the classical case, if the cognitive state of our DM is (operator) $B$
she evaluates the (expected) utility of action $A$ as $\mathbf{Tr}(BA)$%
.\medskip

\textbf{Examples of states.}

1. $S_{0}=E/\dim (H)$, where $E$ denotes the identity operator on $H$. This
is the state of `uniform uncertainty' or the `completely mixed' state. In a
sense, this state is a central point of the set $\mathbf{St}$. The utility
of an action $A$ in this state is equal to $\mathbf{Tr}(A)/\dim (H)$.

2. \emph{Pure states}. Let $e$ be an element of $H$ with length 1 (that is $%
(e,e)=1$). And let $Pr_{e}$ be the orthogonal projector on $e$, that is $%
Pr_{e}(x)=(x,e)e$ for any $x\in H$. $\mathbf{Tr}(Pr_{e})=(e,e)=1$, therefore 
$Pr_{e}$ is a state. Such states are called \emph{pure}. The utility of an
action $A$ in such a state is equal to $(e,Ae)$.

3. Let $S$ be a state, and $U$ a unitary operator (that is $U^{-1}=U^{\ast }$%
), then the operator $USU^{-1}$ is a state as well. Indeed, it is
non-negative, and $\mathbf{Tr}(USU^{-1})=\mathbf{Tr}(U^{-1}US)=\mathbf{Tr}%
(S)=1$. Note that the state $S_{0}$ is invariant under any unitary
conjugation.

4. The set of states \textbf{St} is a convex space. If $S_{1},...,S_{r}$ are
states, and $p_{1},...,p_{r}$ are non-negative real numbers such that $%
p_{1}+...+p_{r}=1$, then the convex combination $p_{1}S_{1}+...+p_{r}S_{r}$
is a state as well. The extreme points of this convex space are exactly the
pure states. Any states can be represented as a convex combination of pure
states; therefore non-pure states are also called mixed states.

As in the classical setting, the utility of an action $A$ is an affine
function ($B\mapsto \mathbf{Tr}(BA)$) on the convex state space \textbf{St}.
\medskip

We understand cognitive states as Receiver's beliefs about the state of some
relevant quantum system.\footnote{%
In our context, the relevant system is the represented world, see \cite%
{Dualm16}.} Receiver chooses an action from some (finite) set of available
actions to maximize her utility. Her optimal action in state $B$ is denoted
as $A^{opt}(B)$. As in the classical case, Sender may want to change
receiver's beliefs with the help of some information structure in order to
induce a choice of action\textbf{\ }that is better for him (than the one
Receiver would choose based on her prior). Below we define more precisely
what we mean by information and its impact on Receiver's belief.
Essentially, this is the main and only difference between the quantum and
the classical problem of persuasion.\medskip

\textit{Information and its impact.} Information is generated by an
information structure (IS) which consists of a measurement device (MD) and
an informational channel (IC). The MD measures the system of interest and
produces an outcome whereas the IC transforms the outcomes of the
measurement device into signals. Receiver is informed about the performance
of the MD, observes the signal and updates her prior belief. Formally, an IS
is defined by three things: the set $S$ of signals, the collection of
probabilities $p_{s}$ for the signals $s\in S$ (which depends on the state $%
B)$ and the collection of associated posteriors $B_{s}$ that is the updated
priors.\textbf{\ }In the present paper which focuses on updating, we have to
distinguish between the two parts of the process (MD and IC). This is
because the MD acts at the `quantum' level (connected with state transition)
whereas the IC is a purely `classical' operation. Most clearly this is
illustrated by the fact that a measurement whose different outcomes are
collected into one and the same signal (which we call `blind' measurement
see below) nevertheless induces a change in Receiver's belief. The
measurement device (MD) is the core of any IS, and we have to describe it in
more detail.\medskip

{We next define a MD for the case the outcomes are directly communicated to
Receiver that is when the signals are equal to the outcomes.\ Thereafter, we
introduce information channels. }The simplest MD consists of a single
projective (or direct) measurement. \medskip

\textbf{Definition.} A \emph{direct (or projective) measurement} is given by
an orthogonal decomposition of the unit (ODU)\textbf{\ }$H=\oplus _{i\in
I}H_{i}$ of Hilbert space $H$ (or equivalently, by a family $P=(P_{i},\ i\in
I)$ of projectors, such that $\sum_{i}P_{i}=E$). The set of outcome of this
measurement is the set $I$; the probability $p_{i}$ of outcome $i$ in state $%
B$ is given by \emph{Born rule} as $p_{i}=\mathbf{Tr}(BP_{i})=\mathbf{Tr}%
(P_{i}BP_{i})$. {\ Upon obtaining a signal (outcome) $i$ the prior $B$
transits into the updated state $B_{i}$ given by} \emph{L\"{u}ders rule} $%
B_{i}=P_{i}BP_{i}/p_{i}$.\footnote{%
A behavioral justification for this rule is provided in \cite{danalmver}.
And in \cite{BelCas} L\"{u}ders rule is shown to be a direct generalization
of Bayes rule.\ } \medskip

Note that $p_{i}\geq 0$ due to Proposition 1. Moreover, $\sum_{i}p_{i}=1$;
this is a simple consequence of the equality $\sum_{i}P_{i}=E$. L\"{u}ders
rule expresses {the quantum nature of our setting. In the classical world, a
measurement that generates new information changes beliefs but not the state
itself. In the quantum world, measurements can essentially change the state
of the measured system and, of course, the belief about the state. As a
consequence }the expected state $B^{ex}=\sum_{i}p_{i}B_{i}=%
\sum_{i}P_{i}BP_{i}$ is generally not equal to the prior $B$. This is in
contradistinction with the classical case where $B^{ex}=B$, a property
called \emph{Bayesian plausibility}. Note also that projective measurements
are endowed with the property of \emph{repeatability}: if we obtain an
outcome $i$ then a repeated measurement (with the same MD) gives the same
outcome $i$. When all the projectors $P_{i}$ are one-dimensional (or pure
states), we have a \textit{complete} measurement; in this case $B_{i}=P_{i}$
independently of $B$.

Schematically, a direct measurement $\mathcal{M}$ can be represented as a
tree where the branches correspond to the possible outcomes. {Which outcome
obtains} as the result of performing the measurement depends of course on
the state of the system. An important feature of the quantum situation is
that this relation is probabilistic: the state of the system only defines
the probabilities for the different outcomes. We represent MD $\mathcal{M}$
as follows

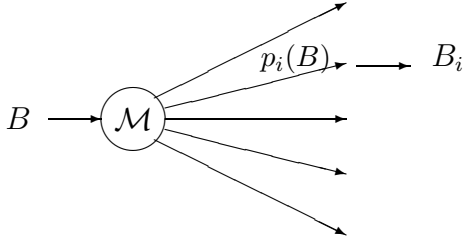
\unitlength=1.4mm \special{em:linewidth 0.4pt} \linethickness{0.4pt} 
\begin{picture}(82.00,34.00)(15,3)
\put(60.00,20.00){\circle{6.00}}
\put(63.00,20.00){\vector(1,0){17.00}}
\put(63.00,21.00){\vector(4,1){17.00}}
\put(63.00,19.00){\vector(4,-1){17.00}}
\put(62.00,18.00){\vector(2,-1){18.00}}
\put(62.00,22.00){\vector(2,1){18.00}}
\put(60.00,20.00){\makebox(0,0)[cc]{$\mathcal{M}$}}
\put(52,20){\vector(1,0){5}}
\put(48.00,19.00){\text{$B$}}
\put(72,25.00){\small{\text{$p_i(B)$}}}
\put(88,25.00){\text{$B_i$}}
\put(81,25){\vector(1,0){5}}
\end{picture}

\noindent where $B$ is the initial state of quantum system, $p_{i}(B)$ the
probability of outcome $i$, and the $B_{i}$ are the resulting (updated)
states.\medskip

Above we addressed single measurements. However, Sender can also use a
compound measurement consisting of a sequences of measurement devices (of
course, the order in which the measurements are performed is important).
Moreover, devices applied consecutively can be conditional, that is they can
depend on the outcome of the previous one. The diagram below illustrates
that point

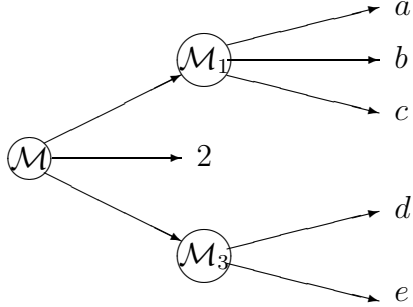
\unitlength=1.00mm \special{em:linewidth 0.4pt} \linethickness{0.4pt} 
\begin{picture}(105.00,45.00)
\put(60.00,20.00){\circle{6.00}}
\put(63.00,20.00){\vector(1,0){17.00}}
\put(62.00,18.00){\vector(2,-1){18.00}}
\put(62.00,22.00){\vector(2,1){18.00}}
\put(60.00,20.00){\makebox(0,0)[cc]{$\mathcal{M}$}}
\put(83.00,33.00){\circle{7.00}}
\put(83.00,33.00){\makebox(0,0)[cc]{$\mathcal{M}_1$}}
\put(86.00,33.00){\vector(1,0){20.00}}
\put(108,39){\text{$a$}}
\put(108,32){\text{$b$}}
\put(108,25){\text{$c$}}
\put(86.00,35.00){\vector(4,1){20.00}}
\put(86.00,31.00){\vector(4,-1){20.00}}
\put(83.00,7.00){\circle{7.00}}
\put(83.00,7.00){\makebox(0,0)[cc]{$\mathcal{M}_3$}}
\put(86.00,8.00){\vector(4,1){20.00}}
\put(86.00,6.00){\vector(4,-1){20.00}}
\put(108,12){\text{$d$}}
\put(108,1){\text{$e$}}
\put(82,19){\text{2}}
\end{picture}\medskip

\noindent Here measurement $\mathcal{M}_{1}$ is applied when the outcome 1
of measurement $\mathcal{M}$ occurs, after outcome 2 the measurement stops,
and after outcome 3, measurement $\mathcal{M}_{3}$ is applied. The set of
outcomes of this compound MD is $\{a,b,c,2,d,e\}$. The probability $p_{a}$
is equal $p_{1}(B)p_{a}(B_{1})$ whereas $p_{d}=p_{3}(B)p_{d}(B_{3})$. Upon
receiving signal $a$ Receiver updates her beliefs to $B_{a}=\frac{%
P_{a}(P_{1}BP_{1})P_{a}}{p_{a}}$.

Let us give a formal definition of a two-stage compound measurement. Suppose
that the first measurement is given by an ODU $(P_{i},i\in I)$, and that for
every $i$ there is a `second' measurement $M_{i}$ given by an ODU ($%
Q_{j},j\in J_{i})$. This compound MD has the set of outcomes $\cup _{i\in
I}(\{i\} \times J_{i})$. The probability of outcome $(i,j)$ (where $j\in
J_{i} $) is $p(i,j)=\mathbf{Tr}(Q_{j}(P_{i}BP_{i})Q_{j})=p_{j}(B_{i})$.
After obtaining outcome $(i,j)$ the system transits into state $%
B_{i,j}=Q_{j}P_{i}BP_{i}Q_{j}/p(i,j)$.

In a similar way one can build more complicate trees of MDs. In our quantum
context, we need to consider compound MDs because the composition of two (or
more) non-commuting\textbf{\ } measurements\textbf{\ }is not a projective
measurement.\footnote{%
Compound measurements belong to a general category of quantum measurements
called POVM. However, nothing can be said in general about where such
measurements takes the state. In order to address the issue of state
transition and updating one needs to consider the details of the procedure.}

We now introduce the possibility that the outcomes of the measurement are
not directly communicated. After the performance of a measurement (possibly
compound), information about the outcome is transmitted to Receiver through
an information channel which transforms the outcomes of the MD into
signals.\medskip

\textbf{Definition.} A (random) \emph{informational channel} (IC) is a
mapping $f:I\rightarrow \Delta (S)$, where $S$ is the set of signals. In the
other words, IC is a family $(f_{s},\ s\in S)$ of non-negative functions on
the set $I$ such that $\sum_{s}f_{s}=1_{I}$. The number $f_{s}(i)$ is the
probability for signal $s$ when outcome $i$ was obtained.\textbf{\ }\medskip

\textbf{Definition.} An \emph{information structure} (IS) is a MD (with a
set $I$ of outcomes) together with an IC $f:I\rightarrow \Delta (S)$.
\medskip

Of course, using an informational channel only "garbles" the information
extracted by the measurement device but this can be in Sender's interest.
Indeed, whenever the prior $B$ is a pure state, the posteriors $B_{i}$
following a projective measurement (or a sequence of direct measurements)
are pure states too. But Sender may prefer Receiver to hold a mixed
posterior. A random information channel can be used for the purpose of
creating mixed posteriors.

Let us return to information structures. The probability $p_{s}$ for signal $%
s$ is equal to $p_{s}=\sum_{i}f_{s}(i)p_{i}$. In this case the initial state
(prior) $B$ transits into the posterior $B_{s}$ given by the formula 
\begin{equation}
B_{s}=(\sum_{i}f_{s}(i)p_{i}B_{i})/p_{s}.  \label{post}
\end{equation}%
Upon receiving signal $s$, Receiver understands (using Bayes rule) that the
conditional probability $p(i|s)$ of outcome $i$ is $f_{s}(i)p_{i}/p_{s}$. {%
She also understands that if she knew for sure that outcome }$i$ occurred
her belief would be represented by $B_{i}$. We can call $B_{i}$
`intermediary beliefs'\textbf{. }Therefore the new cognitive state $B_{s}$
is a mixture of the states $B_{i}$ with weights $p(i|s)$, that is $%
B_{s}=\sum_{i}p(i|s)B_{i}=(\sum_{i}f_{s}(i)p_{i}B_{i})/p_{s}$.

If the informational channel $f$ is the identity mapping from $I$ to $I$, we
speak of a \emph{straightforward} information structure. If the
informational channel is degenerated (that is $S$ is a singleton); in other
words, when Receiver gets no information except that the measurement was
performed, we speak of a \emph{blind} measurement. After a blind
measurement, the posterior $B^{\prime }$ is equal to $\sum_{i}p_{i}B_{i}$.
As earlier emphasized the posterior $B^{\prime }$ generally differs from the
prior $B$.\medskip

\emph{The persuasion problem.} Let us return to the persuasion problem, that
is to the task of changing Receiver's belief. Sender chooses an IS and
announces it to Receiver. Thereafter the measurement is performed, the
information channel is applied and the obtained signal is truthfully
reported to Receiver. Receiver observes the signal, updates her beliefs, and
then takes an (optimal given the new belief) action. Suppose that Sender
chooses an IS with a measurement device $\mathcal{M}$ with the set of
outcomes $I$ and the IC $f:I\rightarrow \Delta (S)$. Suppose further that
the prior beliefs of Receiver are represented by a state $B$, and she
observes a signal $s$. In this case she updates her prior $B$ to the
posterior $B_{s}$. She chooses a new optimal action $A^{opt}(B_{s})$. As a
consequence of using this IS, Sender expects to get utility $%
\sum_{s}p_{s}u(A^{opt}(B_{s}))$.\footnote{%
Here we assume that Sender's belief is equal to Receiver's one.}

Since Sender's objective is to maximize his expected utility, the question
of selecting the optimal IS arises. In the quantum context, this problem is
rather difficult due to the high degree of non-linearity. The problem can be
all the more complicated as Sender can choose to use not a single MD, but a
sequence of MDs. In the classical context we know that this possibility has
no value because any sequence of measurements can be merged into a single
one. This is generally not the case in the quantum context where opting for
a sequence of measurements can significantly increase Sender's persuasion
power. We already mentioned an important result obtained in \cite{TCS}.
Namely, if Sender is not constrained with respect to the number of
measurements, then starting from any prior he can with near certainty have
Receiver's belief transit into any targeted belief state. Of course this
result has mostly a theoretical value. It shows that, in contrast with the
classical case, there exists in principle no obstacle to persuasion.%
\footnote{%
In classical case Bayesian plausibility constitutes a major obstacle.} In
practice however the value of this result is strongly limited because
Receiver is not likely to accept a large number of attempts to be persuaded.
Moreover performing measurements is likely to be time and resource
consuming. Therefore, is it interesting to investigate what Sender can
achieve with very few measurements.

Below we consider a problem closely related to persuasion that we call the 
\emph{targeting}.

\section{Targeting}

As mentioned above we do not address the question of optimal persuasion. We
confine ourselves to a simpler task. Given a prior state $B$ and a `target'
state $T$, we evaluate the probability for the transition from $B$ into $T$
with one measurement or with a small series of sequential measurements. As
usual we consider the case when Sender knows Receiver's prior except in
Proposition 3 where the result is established for unknown prior. The
solution to this task can give a better understanding on how to solve the
full persuasion problem. The connection between the two problems is most
direct when Receiver chooses between two actions only i.e., `Good' and `Bad'
(from the point of view of Sender). We can denote by $D_{g}$ the subset (in 
\textbf{St}) of `good' states (leading Receiver to choose Sender's preferred
action) and by $D_{b}$ the subset of `bad' ones. We can also assume that the
prior $B$ lies in the domain $D_{b}$, because in the opposite case no
measurement is needed. Then Sender chooses some point $T$ in the `good'
domain $D_{g}$ and looks for measurements which gives him this target $T$
with the largest possible probability $p$. This number and associated
expected utility is a lower bound for what can be achieved in the full
persuasion problem\footnote{%
One can say that with `targeting' the DM only cares about the target while
ignoring all other consequences.}. We leave open the important question as
to how to select the point $T$ in the `good' domain $D_{g}$. Intuitively, $T$
should be as close as possible to the prior $B$.

The transition from prior $B$ to a target $T$ is realized with the help of
an IS. More precisely suppose that we have an information structure
consisting of an MD $\mathcal{M}$ with outcomes in $I$ and an information
channel $f:I\rightarrow \Delta (S)$. We say that a signal $t\in S$ is
\textquotedblleft targeting\textquotedblright \ if $t$\ induces the
transition of Receiver's initial cognitive state $B$ into the target $T$
that is if the posterior $B_{t}$\ is equal to $T$. The probability $p_{t}$
for the signal $t$ is called the \emph{probability of transition}.

In other words, a signal $t\in S$ is targeting if the operator $%
\sum_{i}f_{t}(i)p_{i}B_{i}$ (see (\ref{post})) is proportional to the target
state $T$, $pT=\sum_{i}f_{t}\left( i\right) p_{i}B_{i}$; in this case $%
p=p_{t}=\sum_{i}f_{t}(i)p_{i}$ is the probability of transition. When
Receiver learns signal $t$ she updates her belief $B$ into $T$; the signal $%
t $ comes with probability $p_{t}$. Below we use a simplified notation $%
w(i)=f_{t}(i)$ and call $w(i)$ the targeting weights; of course, $0\leq
w(i)\leq 1$, $pT=\sum_{i}w(i)p_{i}B_{i}$, and $p=\sum_{i}w(i)p_{i}$.

Thus, we shall be dealing with the following problem. Given an initial state
(a prior) $B$ and a final state (target) $T$, we search for an IS (including
one MD or a sequence of MDs) which transforms $B$ into $T$ with the largest
possible probability. We know from (\cite{TCS}) that with a large enough
sequence of measurements, starting from any prior, the probability for that
transition can be made arbitrary close to 1. In the next section we show
that this probability can be made larger than $1/n$ with two measurements
only. To this end we below consider in detail two cases: when only one
measurement is feasible and when a series of two measurements is possible.

\subsection{Transition with one measurement}

It is intuitively clear that if the states $B$ and $T$ are close to each
other then the probability $p$ for the transition $B\mapsto T$ can be made
close to 1. Conversely, in \cite{danalmver} we proved that if $p$ is close
to 1 then the states $B$ and $T$ are close as well. On the other hand, the
probability $p$ can vanish. For example, if $B$ and $T$ are pure and
mutually orthogonal states, one expects the probability for transition to be
equal to 0. Indeed, the straightforward measurement $(T,...)$ gives the
posterior $TBT$ which is proportional to $T$, but the probability for the
transition $B\rightsquigarrow T$ is $\mathbf{Tr}(TB)=0$. Sender could use
more subtle MD to achieve his target but the following general statement is
true.\medskip

\textbf{Proposition 2.} \emph{Suppose that prior $B$ and target $T$ are
orthogonal states (that is $\mathbf{Tr}(BT)=0$). Then no information
structure with a single measurement can transform $B$ in $T$}. \medskip

\textit{Proof}. Let $(P_{i},\ i\in I)$ be a direct measurement, and $%
(w(i),i\in I)$ the corresponding targeting weights. We have $%
pT=\sum_{i}w(i)P_{i}BP_{i}$. Multiplying this equation by $B$, we obtain $%
pBT=\sum_{i}w(i)BP_{i}BP_{i}$. Applying the trace, we get equality $%
0=\sum_{i}w(i)\mathbf{Tr}(BP_{i}BP_{i})=\sum_{i}w(i)\mathbf{Tr}%
(P_{i}BP_{i}P_{i}BP_{i})$. Denoting $A_{i}=P_{i}BP_{i}$ we have $%
0=\sum_{i}w(i)\mathbf{Tr}(A_{i}^{2})$. Each term of this sum is non-negative
(see Proposition 1), from where it follows that $\mathbf{Tr}(A_{i}^{2})=0$
if $w(i)>0$. Proposition 1 implies now that $A_{i}=0$ for such $i$. Hence $%
p=\sum_{i}w(i)\mathbf{Tr}(P_{i}BP_{i})=0$. $\Box $ \medskip

On the contrary, if $B$ and $T$ are reciprocally non-orthogonal states, then
one can transform $B$ into $T$ with an {IS that relies on a single MD}. We
shall not prove this here. Instead of, we consider three particular
cases.\medskip

\textbf{1A.} Suppose that the target $T$ is a pure state and is the
projector on unit vector $t$. Let $e_{1},...,e_{n}$ be an orthonormal basis
of $H$ such that $t=e_{1}$. Let $\mathcal{M}$ be the complete direct
measurement on this base. If we make this measurement and obtain the outcome
1 then the prior $B$ transits into pure state $T$ with probability $p=%
\mathbf{Tr}(BT)>0$.

Thus we have shown that \emph{if states $B$ and $T$ are non-orthogonal and $%
T $ is pure then we can transform $B$ into $T$ with positive probability $p$}%
. Of course, $p$ can be very small if $B$ and $T$ are almost orthogonal. On
the contrary, if $B$ and $T$ are close then $p$ is close to 1. \medskip

\textbf{1B. }Let us consider the case when the prior $B$ is the `uniformly
uncertain state' $S_{0}=E/n$, where $n=\dim H$. We assert that \emph{one can
transform $S_{0} $ into any target state $T$ with probability at least $1/n$%
, and explicitly indicate the corresponding MD.}

Suppose that (in some orthonormal basis $e_{1},...,e_{n}$) operator $T$ has
a diagonal form $diag(a_{1},...,a_{n})$. In the other words, $T$ is the
mixture $a_{1}P_{1}+...+a_{n}P_{n}$, where $P_{i}$ are projectors on $e_{i}$%
, $a_{i}\geq 0$, and $a_{1}+...+a_{n}=1$. As the MD we take the complete
direct measurement associated with the basis $e_{1},...,e_{n}$; as{\
targeting weights,} we take $w(i)=a_{i}$. In this case the posterior $B_{t}$
is equal to $(\sum_{i=1}^{n}a_{i}\mathbf{Tr}(S_{0}P_{i})P_{i})/p_{t}$, where 
$p_{t}=\sum_{i=1}^{n}a_{i}\mathbf{Tr}(S_{0}P_{i})$. But $%
P_{i}S_{0}=P_{i}(E/n)=P_{i}/n$, so that $\mathbf{Tr}(S_{0}P_{i})=1/n$ and $%
p_{t}=\sum_{i=1}^{n}a_{i}/n=1/n$. Therefore, the posterior is equal to $%
\sum_{i}a_{i}P_{i}=T$, and the probability of transition is $1/n$.

Note that we may be able to do better. If $\max (a_{i})<1$ then we can take $%
w(i)=a_{i}/\max (a_{i})$ and obtain the desired transition with probability $%
1/n\cdot \max (a_{i})$.\medskip

\textbf{1C.} Note that in point 1B above the operators $B$ and $T$ are
compatible (commuting). Therefore this situation is essentially classical.
The situation is completely different when the target $T$ is equal to $S_{0}$%
. We show (in contrast with the classical case) that \emph{there exists a
blind measurement which transforms any arbitrary prior $B$ (known to Sender)
into $T=S_{0}$ with probability 1}. Of course, that measurement is
incompatible with $B$.\textbf{\ }The following Lemma provides the foundation
for this assertion. \medskip

\textbf{Lemma.} \emph{Let $p_{1},...,p_{n}$ be non-negative numbers, the sum
of which is equal to 1. Then there exists a symmetric $n\times n$ matrix $%
A=(a_{ij})$ with spectrum $p_{1},...,p_{n}$ and diagonal terms $1/n$ (that
is $a_{ii}=1/n$ for all $i$).}\medskip

The Lemma above is a particular case of Horn's wonderful theorem (see for
example, \cite[Theorem 9.B.2]{MarOl}). \medskip

\textbf{Corollary.} \emph{Let $B$ be a state. Then there exists an
orthonormal basis $e_1,...,e_n$ of the space $H$ such that $(e_i,Be_i)=1/n$
for any $i=1,...,n$.}\medskip

\textit{Proof}. Let $p_{1},...,p_{n}$ be the spectrum of $B$; these numbers
are non-negative and their sum is equal to 1. Let $A$ be a symmetric matrix
as in Lemma. Since $A$ has the same spectrum as $B$ there exists a unitary
(indeed, orthogonal) matrix $U$ such that $B=UAU^{-1}$. Define vectors $%
e_{i} $ ($i=1,...,n$) as $U(1_{i})$, where $1_{i}$ is the $i$-th canonical
basis vector in $\mathbf{R}^{n}$. These vectors $e_{i}$ form an orthonormal
basis, and $%
(e_{i},Be_{i})=(U1_{i},UAU^{-1}U1_{i})=(U1_{i},UA1_{i})=(1_{i},A1_{i})=(a_{ij})=1/n 
$. $\Box $\medskip

Consider now the blind complete direct measurement $\mathcal{M}$, given by
the basis $e_{i}$ from Corollary above. The posterior is the mixture of pure
states $e_{i}$ with weights $(e_{i},Be_{i})=1/n$, that is the completely
mixed state $S_{0}=E/n$. The probability of the transition (as for any blind
MD) is equal to 1.

Thus, we proved the following\medskip

\textbf{Theorem 1.} \emph{Any state $B$ can be transformed with one (blind)
measurement into the completely mixed state $S_{0}$ with probability 1.}%
\medskip

This result is quite surprising because it shows that even in the case of a
single measurement, the quantum setting yields results clearly distinct from
those obtained in the classical case. Indeed in the classical context the
support of the posterior must be in the support of the prior distribution.
What is crucial here is that we use a measurement which is incompatible with
prior $B$. We shall see next that this result which in itself is of little
practical value since $T$ seldom is equal to $S_{0}$ is very useful for
targeting with two measurements. We also note that in contrast with the
classical case the naive measurement $T$ (which can be understood as a
direct question on beliefs) because it changes the state has a power of
persuasion as we illustrate in the example below.

\subsection{The case of two measurements: instigating confusion}

\textbf{2A.} Let us consider the following sequence of two measurements.
First, we transform the prior $B$ into completely mixed state $S_{0}$
applying the blind measurement $\mathcal{M}$ defined above in point \textbf{%
1C}. Thereafter, we apply the measurement in \textbf{1B} above which
transforms $S_{0}$ into the target $T$ with a probability of at least $1/n$.

This proves the following\medskip

\textbf{Theorem 2.} \emph{Let prior $B$ and target $T$ be arbitrary states.
Then there exists a sequence of two measurements transforming $B$ into $T$
with a probability of at least $1/n$.}\medskip

We do not assert that the proposed sequence is optimal rather that it
defines a lower bound. Often the transition is achievable with (much) higher
probability; see Proposition 4 below. \medskip

\textbf{2B.} Note that in order to construct the measurement $\mathcal{M}$
in \textbf{1B} above, we need to know the prior $B$. An interesting question
arises: what is achievable when the prior $B$ is unknown to Sender. Below we
provide two particular results in this direction.\medskip

Suppose first that the target $T$ is the completely mixed state $S_{0}$. 
\emph{Then, we can with probability 1, from any arbitrary unknown prior
state }$B,\ $\emph{transit into T with the help of two measurements.} We
make an arbitrary direct measurement (with a base $(e_{1},...,e_{n})$) as
the first measurement $\mathcal{M}_{0}$. Depending on its result $i=1,...,n$
we make the second (blind) measurement $\mathcal{M}_{i}$, given in Theorem
1, which transforms the state $P_{i}=P_{e_{i}}$ into $T=S_{0}$. \medskip

The second case is more subtle. Namely, suppose that our target $T$ is a
pure state, that is the projector on a normalized vector $t$. Due to the
Corollary of the Lemma above (applied now to operator $T$), there exists an
orthonormal basis $e_{1},...e_{n}$ such that $(e_{i},Te_{i})=1/n$ for any $i$%
. Since $Te_{i}=(e_{i},t)t$, we have $1/n=(e_{i},(e_{i},t)t)=(e_{i},t)^{2}$.
In the other words, in basis $e_{1},...e_{n}$ vector $t$ has the form $(1/%
\sqrt{n},...,1/\sqrt{n})$.

Let $P_{i}$ be the projector on vector $e_{i}$. As the first measurement $%
\mathcal{M}$ we take the straightforward measurement in the basis $%
(e_{1},...,e_{n})$. Applying this measurement, the prior $B$ transits into
posteriors $B_{i}=P_{i}$ with probabilities $p_{i}=Tr(P_{i}B)=(e_{i},Be_{i})$%
.

What concerns the second measurement, we take it as the (naive) measurement $%
(T,...)$. This measurement transits any posterior state $B_{i}=P_{i}$ into
the state $TB_{i}T/Tr(TB_{i})=T$ with probability $%
Tr(TB_{i})=Tr(TP_{i})=(t,e_{i})^{2}=1/n$. In the average, we transit into
the state $T$ with probability $\sum_{i}p_{i}\cdot 1/n=1/n$.

Thus we proved the following \medskip

\textbf{Proposition 3.} \emph{If the target }$T$\emph{\ is a pure state then
there exists a sequence of two measurements (depending only on }$T$\emph{)
which transits any (known or unknown to Sender) prior }$B$\emph{\ into }$T$%
\emph{\ with probability at least }$1/n$\emph{.}\medskip

We conjecture that the assertion in Proposition 3 is true for any arbitrary
target $T$, i.e., pure or mixed. \medskip

\textbf{2C.} Let us consider more in detail the two-dimensional case i.e.,
when the dimension of Hilbert space $H$ is equal to 2. Moreover, we suppose
that the prior $B$ and the target $T$ are pure states, given by normalized
vectors $b$ and $t$. Without loss of generality, we can assume that these
vectors have coordinates $t=(1,0)$ and $b=(\cos (\varphi ),\sin (\varphi ))$.

As the first measurement $\mathcal{M}$ we take the direct measurement with
orthonormal basis $(e_{1},e_{2})$, where $e_{1}=(\cos (\varphi /2),\sin
(\varphi /2))$ and $e_{2}=(\sin (\varphi /2),-\cos (\varphi /2))$. After
this measurement, the prior $b$ transits into the state $e_{1}$ with the
probability $p_{1}=(e_{1},b)^{2}=\cos ^{2}(\varphi /2)$ and into the state $%
e_{2}$ with the (complementary) probability $\sin ^{2}(\varphi /2)$.

Under an impact of the second (naive) measurement $(T,E-T)$ the state $e_{1}$
transits into $t$ with probability $(e_{1},t)^{2}=\cos ^{2}(\varphi /2)$
whereas the state $e_{2}$ transits into $t$ with probability $%
(e_{2},t)^{2}=\sin ^{2}(\varphi /2)$. As the result, with this sequence of
two measurements, the prior $b$ transits into the target $t$ with probability%
\textbf{\ $\cos ^{4}(\varphi /2)+\sin ^{4}(\varphi /2)$. }

Since $\cos ^{2}(\varphi /2)=1/2+\cos (\varphi )/2$ and $\sin ^{2}(\varphi
/2)=1/2-\cos (\varphi )/2$, we can rewrite $\cos ^{4}(\varphi /2)+\sin
^{4}(\varphi /2)$ as $1/2+\cos ^{2}(\varphi )/2$. Obviously, this number is
no less than 1/2. Summing up we have the following proposition\medskip

\textbf{Proposition 4.} \emph{When both the prior and the target are pure
states known to Sender we can transform the prior $b$ into the target $t$
with the probability $(1+\cos ^{2}(\varphi ))/2=(1+(b,t)^{2})/2$, where $%
\varphi $ is the angle between the vectors $b$ and $t$.}\medskip

One can easily see that a $n-$dimensional targeting problem essentially
boils down to the two dimensional case when both the prior and the target
are pure states known to Sender. Indeed, the vectors $b$ and $t$ lie in a
two-dimensional subspace of $H$, and we can consider that they have the form 
$t=(1,0,0,...,0)$ and $b=(\cos (\varphi ),\sin (\varphi ),0,...,0)$.%
\footnote{%
The angle $\varphi $ is a measure of the distance between the prior and the
target state.} As the first measurement $\mathcal{M}$ we take the direct
measurement with orthonormal basis $(e_{1},e_{2},...e_{n})$, where $%
e_{1}=(\cos (\varphi /2),\sin (\varphi /2),0...,0)$, $e_{2}=(\sin (\varphi
/2),-\cos (\varphi /2),0,..,0)$, $e_{3}=(0,0,1,0,...,0)$, $e_{4}=\left(
0,0,0,1,0,...,0\right) $ and so on. Further, the reasoning is the same as
above.

\subsection{ Illustration: Persuading a MP to vote No}

We now return to the story about the MP's decision to support or not a law
that introduces a state of emergency to combat terrorism (see Introduction).
We below show how the theory developed above can be used to analyze the
activist's successful persuasion of the MP.

Let $H$ be a two-dimensional Hilbert space with an orthonormal basis $%
e_{1},e_{2}$; the corresponding projectors are $P_{1}=%
\begin{pmatrix}
1 & 0 \\ 
0 & 0%
\end{pmatrix}%
$ and $P_{2}=%
\begin{pmatrix}
0 & 0 \\ 
0 & 1%
\end{pmatrix}%
$. The MP (Receiver) has to choose between two actions: $Yes$ and $No$. We
represent the action $No$ with Hermitian operator $N=%
\begin{pmatrix}
1 & 0 \\ 
0 & -2%
\end{pmatrix}%
$. The utility of this action in the state $P_{1}$ is equal to 1, and is
equal to -2 in the state $P_{2}$. The action $Yes$ is given by Hermitian
operator $Y=%
\begin{pmatrix}
-1 & 0 \\ 
0 & 1%
\end{pmatrix}%
$. If our MP has belief given by $B=%
\begin{pmatrix}
a & b \\ 
b & 1-a%
\end{pmatrix}%
$, then her expected utility of action $No$ is $\mathbf{Tr}(NB)=3a-2$
whereas the expected utility of action $Yes$ is $\mathbf{Tr}(YB)=1-2a$. If $%
a<3/5$ our MP votes $Yes$; if $a\geq 3/5$ the MP votes for $No$.

We assume further that our MP holds initial belief (prior) $B=%
\begin{pmatrix}
1/5 & 2/5 \\ 
2/5 & 4/5%
\end{pmatrix}%
$. With this belief MP votes $Yes$.

The activist (Sender) receives utility 1 when succeeding in rallying an MP
to the $NO$ vote whatever the true level of threat and utility 0 otherwise.
With MP's prior belief, the activist gets the utility 0. Can he persuade
Receiver to vote $NO$ by selecting an appropriate\textbf{\ }information
structure? We would like to illustrate our results in Theorem 2 and
Proposition 4 with this example as well as some issues pertaining to the
choice of the target state. \medskip

1) To begin with, suppose that the activist naively asks MP whether she
believes the threat is moderate (that is $e_{1}$) or severe (that is $e_{2}$%
). In the other words, he makes the direct measurement $(e_{1},e_{2})$.
Already this simple and naive question changes the belief of our MP: with
probability $p_{1}=\mathbf{Tr}(BP_{1})=4/5$ her posterior becomes $P_{1}$,
and with probability 1/5 her posterior becomes $P_{2}$. So even this simple
question increases the (expected) utility of the activist up to 1/5.

Next we show\textbf{\ }that with two measurements Sender can do much
better.\medskip

2) Let us consider another perspective on the law which we call `the quality
of public decision-making' (cf. EU decision in the Introduction). A state of
emergency gives extended new powers to public officials. This quality
property is "tested" by the following direct measurement $(Q_{1},Q_{2})$
with\ two possible outcomes $Q_{1}=%
\begin{pmatrix}
1/2 & 1/2 \\ 
1/2 & 1/2%
\end{pmatrix}%
$ and $Q_{2}=%
\begin{pmatrix}
1/2 & -1/2 \\ 
-1/2 & 1/2%
\end{pmatrix}%
$. Such a measurement corresponds to a basis of the state space (of the
mental representation of the issue) different from the basis $%
(e_{1},e_{2}),\ $a$\ 45^{\circ }$ rotation of $\left( e_{1},e_{2}\right) $.
This means that $\left( P_{1},P_{2}\right) $ and $(Q_{1},Q_{2})$ are two
non-commuting measurements. Or equivalently $(P_{1},P_{2})$ and $%
(Q_{1},Q_{2})$ correspond to properties of the system that are incompatible
in the mind of Receiver. She can think in terms of either one of the two
perspectives but she cannot synthesize (combine in a\ stable way) pieces of
information from the two perspectives.

Assume now that Sender brings up a discussion about the quality perspective
and performs measurement $(Q_{1},Q_{2})$ that is Receiver learns whether or
not the EU regulates cucumbers. With some probability $p$ the belief $B$
transits into $Q_{1}$ (and with the complementary probability into $Q_{2})$.
Thereafter, the activist again asks her whether she believes the threat is
moderate or severe, i.e. performs measurement $\left( P_{1},P_{2}\right) $.
If the posterior $B^{\prime }$ is equal to $Q_{1}$ then with probability $%
\mathbf{Tr}\left( Q_{1}P_{1}\right) =1/2$ the new cognitive state becomes
the target $P_{1}$. Similarly if the posterior is $B^{\prime \prime }=Q_{2}$
with probability $\mathbf{Tr}\left( Q_{2}P_{1}\right) =1/2$ it transits into
the target $P_{1}$. In the state of belief $P_{1}$ our MP votes NO. So with
a probability 1/2 the target is reached and Sender gets an expected utility
of 1/2 (instead of 0 without a persuasion or 1/5 in the case of a direct
question).\medskip

3) Above we considered the case when the target was taken to be $P_{1}$. But
we can take as a target another state $T=%
\begin{pmatrix}
3/5 & \sqrt{6}/5 \\ 
\sqrt{6}/5 & 2/5%
\end{pmatrix}%
$. This target (as a belief state of our MP) also induces action $NO$. As
asserted in Proposition 4, Sender can transit into this target state with
probability $1/2+(b,t)^{2}/2$. In our case $b=(1/\sqrt{5},2/\sqrt{5})$ and $%
t=(\sqrt{3}/\sqrt{5},\sqrt{2}/\sqrt{5})$, so that $(b,t)=\sqrt{3}/5+\sqrt{8}%
/5$, this probability is equal to 0.916! \medskip

4) But a little (quantum) marvel awaits us ahead. We shall show that with
the help of one (blind) measurement Sender persuade MP to vote No with
certainty.

For this purpose, Sender selects the blind measurement $(Q_{1},Q_{2})$,
where $Q_{1}=%
\begin{pmatrix}
\frac{1+c}{2} & c \\ 
c & \frac{1-c}{2}%
\end{pmatrix}%
$ and $c=1/\sqrt{5}$. Correspondingly, $Q_{2}=%
\begin{pmatrix}
\frac{1-c}{2} & -c \\ 
-c & \frac{1+c}{2}%
\end{pmatrix}%
$. The probabilities are $p_{1}=(1+c)/2$ and $p_{2}=(1-c)/2$. The posterior
is mixed state $B^{ex}=p_{1}Q_{1}+p_{2}Q_{2}=%
\begin{pmatrix}
3/5 & 1/10 \\ 
1/10 & 2/5%
\end{pmatrix}%
$ and MP chooses to vote No with probability 1!\medskip

In our example, we assume that the danger perspective (threat level) which
appeals to geopolitical arguments and emotional ones (fear) and the
perspective on the quality of decision-making in public administrations
which appeals to arguments about bureaucratic senselessness and the
intrusiveness of the state, are two incompatible perspectives in the MP's
mind. She can think in either perspective but find it difficult to deal with
them simultaneously although they are both relevant to the issue at stake.

The example illustrates how Sender can exploit the quantum indeterminacy of
the cognitive state (expressed in the incompatibility of the two
perspectives) to persuade our (quantum-like) decision-maker. By performing a
measurement on an incompatible perspective, the cognitive state is modified
such that beliefs with respect to the severeness of the threat are updated
so that Receiver prefers to vote NO with probability 1.

\section{Concluding remarks}

The theory of Bayesian persuasion establishes that it is often possible to
influence a rational decision-maker by selecting a suitable information
structure and making the corresponding measurement. However, in the
classical uncertainty environment the persuasion power is strongly limited
by Bayesian plausibility. The reality of persuasion or manipulation seems
however far more extended. Therefore, we have here extended it to the
non-classical (quantum) uncertainty environment. This corresponds to
investigating persuasion with a quantum cognitive approach - an avenue of
research that has experienced rapid growth under the latest 20 years. In a
recent paper \cite{TCS}, we establish that provided Sender can make as many
measurement as he wishes, full persuasion is always feasible. Sender can
bring Receiver to believe whatever he wants. This theoretical result
suggests that quantum persuasion is powerful indeed compared with Bayesian
persuasion. But no one expects Receiver to accept attempts to persuasion
under an indefinite time. This paper investigates what can be achieved with
short sequences of measurements.

While Bayesian plausibility imposes a constraint, it also allows to simplify
the persuasion problem so as to allow characterizing an optimal persuasion
policy. In the non-classical context, no such simplification is available
and the issue of optimality cannot be directly addressed. A simpler task
that we call targeting is investigated and its results can be viewed as a
lower bound for what can be achieved in a full persuasion problem. Our
results show that even with a short sequence of two measurements, targeting
is quite powerful. In particular when the prior is known (and is a pure
state), whatever the initial belief, any target belief can be reached at
least half of the time. Most interesting is to understand how this is made
possible. The first step amounts to creating what we would like to call
confusion (formally uniform uncertainty). This is possible precisely because
any quantum belief state (even pure) can always be \textquotedblleft
broken\textquotedblright \ by using a non-commuting measurement. In quantum
cognition this expresses the fact that people have difficulties to process
different kinds of information into a single and stable representation of
the world. Our analysis suggests that this very feature makes a person
easily influenceable. And that a person can be persuaded more effectively
using well-designed distraction rather than by arguments of informational
value to the decision. Distraction occurs when the mind is turned toward a
perspective that Receiver finds hard to combine with her current
perspective. It is a common experience that sellers use this tactic
intuitively sensing that it is quite efficient. We have here provided an
argument based on purely informational considerations that justifies a
distraction policy in influence seeking activities.

\end{document}